\documentstyle[titlepage,12pt]{article}
\newcommand{\preprint}[2]{\large
\begin{center}
\hspace*{8cm} Preprint  \, EFUAZ\,  94-{#1}
\hspace*{8cm}  {#2} 1994
\end{center}}

\def\@makefnmark{\mbox{$^\@thefnmark$}}

\long\def\@makefntext#1{\noindent$^{\@thefnmark}$#1}

\def\openone{\leavevmode\hbox{\small1\kern-3.3pt\normalsize1}}

\begin{document}
\thispagestyle{empty}

\title{\vspace*{-3cm}\preprint {03}{April}\vspace*{2cm}
{\large {\bf  THE DIRAC-DOWKER OSCILLATOR}}\thanks{Submitted to {\it Physics
Letters A}}}\vskip6mm

\author{
{\bf  Valeri V. Dvoeglazov}\thanks{On leave of absence from \,
{\it Dept. Theor. \& Nucl. Phys., Saratov State University, Astrakhanskaya
str., 83, Saratov RUSSIA}}\,\,\,\thanks{Email: {\it valeri@bufa.reduaz.mx,
dvoeglazov@main1.jinr.dubna.su}} \\
{\it Escuela de F\'{\i}sica, Universidad Aut\'onoma
de Zacatecas}\\
{\it Ciudad Universitaria, Antonio Doval\'{\i} Jaime,\, s/n}\\
{\it  Zacatecas, ZAC., M\'exico\,  98068}}
\date{\empty}
\maketitle

\begin{abstract}
The oscillator-like interaction is introduced in  the equation for  the
particle of arbitrary spin, given by Dirac and re-written to a matrix form by
Dowker.
\end{abstract}
\maketitle

\newpage
\setcounter{page}{1}

Recently, the conception of relativistic oscillators~\cite{1} got developed
significantly in
connection with the papers of Moshinsky~\cite{Mosh}. The case of spin $j=1/2$,
the Dirac
oscillator, ref. [2a],
as well as the cases of $j=0$, the Klein-Gordon oscillator, ref.~\cite{Bruc},
and  $j=1$, the Duffin-Kemmer oscillator, ref.~\cite{Deb}, have been
investigated. In this letter we try to
generalize the notion of  relativistic oscillator with intrinsic spin structure
to the case of
arbitrary spin.

We begin  from the equation for any spin given by Dirac~\cite{Dirac}, see also
ref.~\cite{Fierz}, in the form
written down in  ref.~\cite[p.154]{Corson} by Corson, $c=\hbar=1$,
\begin{eqnarray}\label{eq:DD}
\cases{\partial^{\dot A B}v_B (k+{1\over 2}) \psi (k+{1\over 2}, l-{1\over
2})-m \left (\frac{2k+1}{2l}\right )^{1/2}v^{\dot A}(l) \psi (k,l)=0,&\cr
&\cr
\partial_{\dot A B}v^{\dot A} (l) \psi (k, l)+m \left (\frac{2l}{2k+1}\right
)^{1/2} v_{B} (k+{1\over 2}) \psi (k+{1\over 2},l-{1\over 2})=0.&}
\end{eqnarray}
Here we  use Corson's notation,
 $v^A$ and $v^{\dot A}$ are the rectangular spinor-matrices of  $2k$ rows and
$2k+1$ columns (see, e. g., section 17b of ref.~\cite{Corson} for details).
The wave function $\psi (k,l)$ belongs to the $(k,l)$ representation of the
homogeneous Lorentz group.
The choice $l=1/2$ and $k=j-1/2$, $j$ is the spin of a particle, permits one to
reduce a number of
subsidiary conditions. Namely, we have the only condition in this
case~\cite{7a,Dowker2}:
\begin{equation}
\partial_{\dot A B} v^B (j-{1\over 2})v^{\dot A} ({1\over 2})\psi (j-{1\over
2},{1\over 2})=0,
\end{equation}
which follows from (1b).
Moreover, the equations (\ref{eq:DD}) are shown  by
Dowker~\cite{Dowker1}-\cite{Dowker4} to recast to the
matrix form which is similar to the well-known Dirac equation for $j=1/2$
particle
\begin{eqnarray}\label{eq:DDo}
\cases{\alpha^\mu \partial_\mu \Phi =  m\Upsilon,&\cr
\bar\alpha^\mu \partial_\mu \Upsilon = -m \Phi .&}
\end{eqnarray}
The $4j$- component function $\Phi$ could be identified with the wave function
in $(j,0)\oplus (j-1,0)$ representation. Then,  $\Upsilon$, which also has $4j$
components, is written down
\begin{eqnarray}
\Upsilon = (-1)^{2j} (2j)^{-{1\over 2}}\left (\matrix{
v_{\dot A} (j-{1\over 2})\otimes v^{\dot A} ({1\over 2})\cr
u_{\dot A} (j) \otimes v^{\dot A} ({1\over 2})\cr
}\right ) \psi (j-{1\over 2}, {1\over 2}).
\end{eqnarray}
and it belongs to $(j-1/2,1/2)$ representation.
Four matrices $\alpha^\mu$ and  four matrices $\bar\alpha^\mu=\alpha_\mu$,
where $\mu=0,1,2,3$, \,   obey  the anticommutation
relations of Pauli matrices
\begin{equation}\label{eq:ac}
\bar \alpha^{(\mu} \alpha^{\nu)}=g^{\mu\nu},
\end{equation}
however, they have the dimension $4j\otimes 4j$.
This set of matrices has been investigated in details in ref.~\cite{Dowker1},
and  $\alpha^\mu$  was proved there to satisfy all the algebraic relations of
the Pauli matrices except for completeness.

Defining $p_\mu=-i\partial_\mu$ and the analogs of $\gamma$-  matrices as
following:
\begin{eqnarray}\label{eq:ga}
\gamma^\mu = \pmatrix{
0 & -i\bar\alpha^\mu \cr
i\alpha^\mu & 0 \cr
}
\end{eqnarray}
the set of equations (\ref{eq:DDo}) is written down to the form of the Dirac
equation
\begin{equation}\label{eq:Di}
\left (p_\mu \gamma^\mu - m\right ) \left (\matrix{
\Phi\cr
\Upsilon\cr
}\right ) =0.
\end{equation}
However, let us not forget that $\Phi$ and $\Upsilon$ are 2-spinors only in the
case of $j=1/2$.

As mentioned in, e. g., ref.~\cite[p.33,124]{Corson}, in the case of spin
$j=1/2$ the set of $\gamma$- matrices in representation (\ref{eq:ga})
\begin{eqnarray}
\gamma^0 &=& \pmatrix{
0 & -i \openone_{2\otimes 2}\cr
i \openone_{2\otimes 2} & 0 \cr},
 \hspace*{1cm} \gamma^1=\pmatrix{
0 & i\tau^1 \cr
i\tau^1 & 0 \cr
},\nonumber\\
& &\nonumber \\
\gamma^2 &=& \pmatrix{
0 & i\tau^2  \cr
i \tau^2 & 0 \cr
}, \hspace*{1cm} \gamma^3=\pmatrix{
0 & i\tau^3 \cr
i\tau^3 & 0 \cr
}
\end{eqnarray}
is defined up to the unitary transformation and  Eq. (\ref{eq:Di}) could be
recast to the Hamiltonian form given by Dirac (with $\alpha_k$ and $\beta$
matrices) by means of the unitary matrix. It is easy to carry out the same
procedure ($\alpha^k= {\cal S} \gamma^0\gamma^k {\cal S}^{-1}$ and $\beta={\cal
S} \gamma^0 {\cal S}^{-1}$) for $\gamma$ matrices, Eq. (\ref{eq:ga}), and
functions of arbitrary spin ($\Psi= {\cal S}^{-1}\cdot column (\phi ,
\upsilon)$).  For our aims it is convenient to choose the unitary matrix as
following:
\begin{eqnarray}
{\cal S} ={1\over \sqrt{2}}\pmatrix{
\openone_{4j\otimes 4j} & i \openone_{4j\otimes 4j} \cr
i \openone_{4j\otimes 4j} & \openone_{4j\otimes 4j} \cr
}.
\end{eqnarray}
 After standard substitution $\vec p \rightarrow \vec p -im\omega \gamma^0 \vec
r$ we obtain
\begin{eqnarray}
\cases{E\phi = -i\left [ \alpha_0 (\vec \alpha\vec p) +im\omega (\vec \alpha
\vec r)\right ]\upsilon +m\alpha_0\phi, & $$\cr
E\upsilon = i \left [ \alpha_0(\vec \alpha \vec p) -im\omega (\vec \alpha \vec
r)\right ]\phi -m \alpha_0\upsilon. & $$}
\end{eqnarray}
Since it follows from the anticommutation relations (\ref{eq:ac}) that
$\alpha_i \alpha_0=\alpha_0 \alpha_i$ (the signature is taken to be $(+\  -\
-\  -)$),  we have the equations
which coincide with Eq. (8)  of ref. [2a] or  Eqs. (3.6) and (3.12) of ref.
[2e] except for $ \tau_\mu \rightarrow  \alpha_\mu$, i. e. their explicit
forms,
\begin{eqnarray}
(E^2 -m ^2)\phi &=& \left [ \vec p^{\,2}+ m\omega \vec r^{\,2} -3\alpha_0
m\omega -  2m\omega \alpha_0 \Sigma^{ij} r^i  \bigtriangledown_j\right ]\phi\\
(E^2 +m ^2)\upsilon &=& \left [\vec p^{\,2}+ m\omega \vec r^{\,2} +3\alpha_0
m\omega +  2m\omega \alpha_0 \Sigma^{ij} r^i  \bigtriangledown_j \right
]\upsilon.
\end{eqnarray}
where ${\Sigma}^{ij}={1\over 2}\alpha^{[ i}\bar\alpha^{j ]} $. As shown
in~\cite{Dowker1} it is always possible to pass to the representation of
$\alpha^\mu$ matrices with $\alpha^0$ the unit matrix.

Thus,  we  convinced ourselves that we obtained the same oscillator-like
interaction and the similar spectrum as for the case of $j=1/2$ particles in
[2a].

\medskip
The author greatly appreciate valuable discussions with Profs. J. Beckers,
N. Debergh and A. del Sol Mesa. I am also grateful to Prof. J. S. Dowker \, for
sending me the reprints of  his  papers.

\end{document}